\documentclass[doc]{article}
\usepackage[utf8]{inputenc}
\usepackage[english]{babel}
\usepackage{mathptmx}
\usepackage{amsmath}
\usepackage{graphicx}

\begin{document}

\setcounter{secnumdepth}{2}

\title{Modelling Student Behavior using Granular Large Scale Action Data from a MOOC}


\author{
  Steven Tang\\
  UC Berkeley\\
  \texttt{steventang@berkeley.edu}
  \and
  Joshua C. Peterson\\
  UC Berkeley\\
  \texttt{peterson.c.joshua@gmail.com}
  \and
  Zachary A. Pardos\\
  UC Berkeley\\
  \texttt{pardos@berkeley.edu }
}


\maketitle
\begin{abstract}
 Digital learning environments generate a precise record of the actions learners take as they interact with learning materials and complete exercises towards comprehension. With this high quantity of sequential data comes the potential to apply time series models to learn about underlying behavioral patterns and trends that characterize successful learning based on the granular record of student actions. There exist several methods for looking at longitudinal, sequential data like those recorded from learning environments. In the field of language modelling, traditional n-gram techniques and modern recurrent neural network (RNN) approaches have been applied to algorithmically find structure in language and predict the next word given the previous words in the sentence or paragraph as input. In this paper, we draw an analogy to this work by treating student sequences of resource views and interactions in a MOOC as the inputs and predicting students' next interaction as outputs. In this study, we train only on students who received a certificate of completion. In doing so, the model could potentially be used for recommendation of sequences eventually leading to success, as opposed to perpetuating unproductive behavior. Given that the MOOC used in our study had over 3,500 unique resources, predicting the exact resource that a student will interact with next might appear to be a difficult classification problem. We find that simply following the syllabus (built-in structure of the course) gives on average 23\% accuracy in making this prediction, followed by the n-gram method with 70.4\%, and RNN based methods with 72.2\%. This research lays the ground work for recommendation in a MOOC and other digital learning environments where high volumes of sequential data exist.  

\end{abstract}

\section{Introduction}
Today's digital world is marked with personalization based on massive logs of user actions. In the field of education, there continues to be research towards personalized and automated tutors that can tailor learning suggestions and outcomes to individual users based on the (often latent) traits of the user. In recent years, massive sources of student-generated learning actions have been collected by online learning environments, such as in Massive Open Online Courses (MOOCs). Utilizing such granular and massive data has been an ongoing research problem. In this paper, we seek to contribute to the growing body of research that aims to utilize large sources of student-created data towards the ability to personalize learning pathways to make learning as accessible, robust, and efficient as desired. We depart from previous research objectives concerned primarily with assessing student knowledge, which is the primariy objective of knowledge tracing, and pivot to modeling of student behavior. We seek to consider all actions of students in a MOOC such as viewing lecture videos or replying to forum posts and attempt to predict their next action. Such an approach makes use of the granular, non assesment data collected in MOOCs and has potential to provide a wide range of recommendations for students looking for navigational guidance in MOOCs.

As an example of why digitized actions can be useful, when a student in a MOOC is struggling on a quiz, researchers and educators can investigate log data to track down exactly which actions the student takes next and whether those actions eventually lead to a successful learning result in the future. Utilizing such data across tens of thousands of students engaged in MOOCs, we ask whether trends of successful navigation through MOOCs can be uncovered by modelling the behavior of students who were ultimately succesful in the course. Capturing the trends that successful students take through MOOCs can enable the development of automated recommendation systems so that struggling students can be given meaningful and effective recommendations to optimize their time spent trying to succeed. For this task, we utilize n-gram and recurrent neural network models that have been traditionally successful when applied to other generative and sequential tasks.

This paper specifically analyzes how well such models can predict the next action given a context of previous actions the student has taken. The purpose of such analysis would be to eventually create a system whereby an automated recommender could query the model to provide meaningful guidance on what action the student can take next. The next action in many cases may be the next resource perscribed by the course but in other cases it may be a recommendation to consult a resource that is back in a previous lesson or enrichment material that is buried in a corner of the courseware unknown to the student. These models we are training are known as generative, in that they can be used to generate what action could come next given a prior context of what actions the student has already taken. Actions can include things such as opening a lecture video, pausing the lecture video, answering a quiz question, or navigating and replying to a forum post. This research serves as a foundation for applying sequential, generative models towards creating personalized recommenders in MOOCs with potential applications to other educational contexts with sequential data.

\section{Related Work}

In this work, we seek to create a generative model of student learning, such that the model can produce or suggest a potential new action for the student to complete based on their past work. This type of generative task has been explored in other fields.

A simple but powerful model used in natural language processing (NLP) is the \textit{n}-gram model \cite{brown1992class}, where a probability distribution is learned over every possible sequence of \textit{n} terms from the training set. More recently, recurrent neural networks (RNNs) have been used to perform next-word prediction \cite{mikolov2010recurrent}, where previously seen words are subsumed into a high dimensional continuous latent state. This type of approach has the potential to utilize contexts of arbitrary length.

Other work has been done to map assignments, student ability, lesson gains and pre-requisites all onto the same dimensional embedding space, where dimensions represent latent skills \cite{DBLP:journals/corr/ReddyLJ16}. This mapping enables the model to produce potential lesson and assignment pathways so that students can have an optimal path for learning. The work in this paper differs by attempting to utilize all sources of student actions, such as forum posts or specific video viewings, and as such requires the use of a different approach.

More broadly, much work has been done to assess the latent knowledge of students through models such as Bayesian Knowledge Tracing (BKT) \cite{corbett1994knowledge}. This type of modelling views the actions of students' as learning opportunities to model student latent knowledge. The work in this paper is tangentially related, in that student knowledge is not explicitly modelled. Instead, the models in this paper focus on predicting what the student will do next; latent student knowledge may play a role in determining what should be done next, but is not explicitly made as part of the model.

Deep Knowledge Tracing \cite{piech2015deep} uses recurrent neural networks to create a continuous latent representation of students based on previously seen assessment results as they navigate online learning environments. This work shows that a deep learning approach can be used to represent student knowledge, with favorable accuracy predictions relative to shallow BKT. Such results, however, are hypothesized to be explained by already existing extensions of BKT \cite{khajah2016deep}. The use of deep learning to approach knowledge tracing still finds useful relationships in the data automatically, but potentially does not find additional representations relative to already proposed extensions to BKT. The work in this paper is related to the use of deep networks to represent students, but differs in that all types of student actions are considered rather than only the use of assessment actions.

Specifically, in this paper we consider using both the \textit{n}-gram approach as well as using a variant of the RNN known as the Long Short-Term Memory (LSTM) architecture \cite{lstm1997}. These two models are chosen as they are both used to model sequences of data and provide a probability distribution of what token should come next. The use of LSTM architectures and similar variants have recently achieved impressive results in a variety fields that involve sequential data, including speech, image, and text analysis \cite{graves2013speech,NIPS2015_5635,Vinyals_2015_CVPR}, in part due to its mutable memory that allows for the capture of long- and short-range dependencies in sequences. Since student learning behavior can be represented as a sequence of actions from a fixed action state space, LSTMs could potentially be used to capture complex patterns that characterize successful learning. In previous work, modelling of student clicksteam data has shown promise with methods such as n-gram models \cite{wen2014identifying}.

\section{Dataset}
The dataset used in this paper came from a Statistics BerkeleyX MOOC from Spring 2013. The MOOC ran for five weeks, with video lectures, homework assignments, discussion forums, and two exams. The original dataset contains 17 million actions from around 31,000 students, where each action represents accessing a particular link in the course (video view, assignment view, problem view etc.). Of the 31,000 students, 8,094 completed enough assignments and scored high enough on the exams to be considered ``certified'' by the instructors of the course. Note that in other MOOC contexts, certification sometimes means that the student paid for a special certification, but that is not the case for this MOOC. The certified students accounted for 11.2 million of the original 17 million actions, with an average of 1,390 actions per certified student. The distinction between certified and non-certified is important for this paper, as we chose to train the generative models only on actions from students who were considered ``certified,'' under the hypothesis that the sequence of actions that certified students take might reasonably approximate a successful pattern of navigation for this MOOC.

Each row in the dataset contained relevant information about the action, such as the exact URL of what the user is accessing, a unique identifier for the user, the exact time the action occurs, and more. For this paper, we do not consider time or other possibly relevant contextual information, but instead focus solely on the resource the student accesses. Actions that occurred fewer than 40 times throughout the entire dataset were removed, as those tended to be discussion posts or user profile visits that were rarely accessed and are unlikely to be applicable to other students navigating through the MOOC. After removing infrequent actions, 3,687 unique actions remained in the model. Thus, all of our models were trained to predict actions taken only from this set.

Although our goal was to keep the data in its most original form as possible, there were some steps we took to extract meaningful and unique actions from the log data. For actions that had an event\_type of ``save\_problem\_check'', we replaced that action with the corresponding object\_name from the data, as this was a better unique identifier for which problem the student was checking. Some of the actions in the log data contained an explicit ``page'' corresponding to exactly what resource they were accessing; if this page was in the row, it was used as the action for that row. Otherwise, the value in event\_type was used.

$15\%$ of the unique resources in our dataset were comprised of transitions to unique courseware content pages (i.e. problem sets). The remaining possible resources ($85\%$) indicated transitions to unique social content pages (i.e. forum threads). However, only $6\%$ of actions taken were related to social content. No individual action frequency surpassed $3\%$ of the entire dataset. 240 ($6.5\%$) of the resource choices accounted for $90\%$ of actions taken by students.

\section{Methodology}
In this section, we detail the architecture of the recurrent neural network and the LSTM extension, which is the model that we hypothesize will perform best at next-action prediction. Other ``shallow'' models, such as the \textit{n}-gram, are described afterwards.

\subsection{Recurrent Neural Networks}
\begin{figure}[!h]
\centering
\includegraphics[width=80mm]{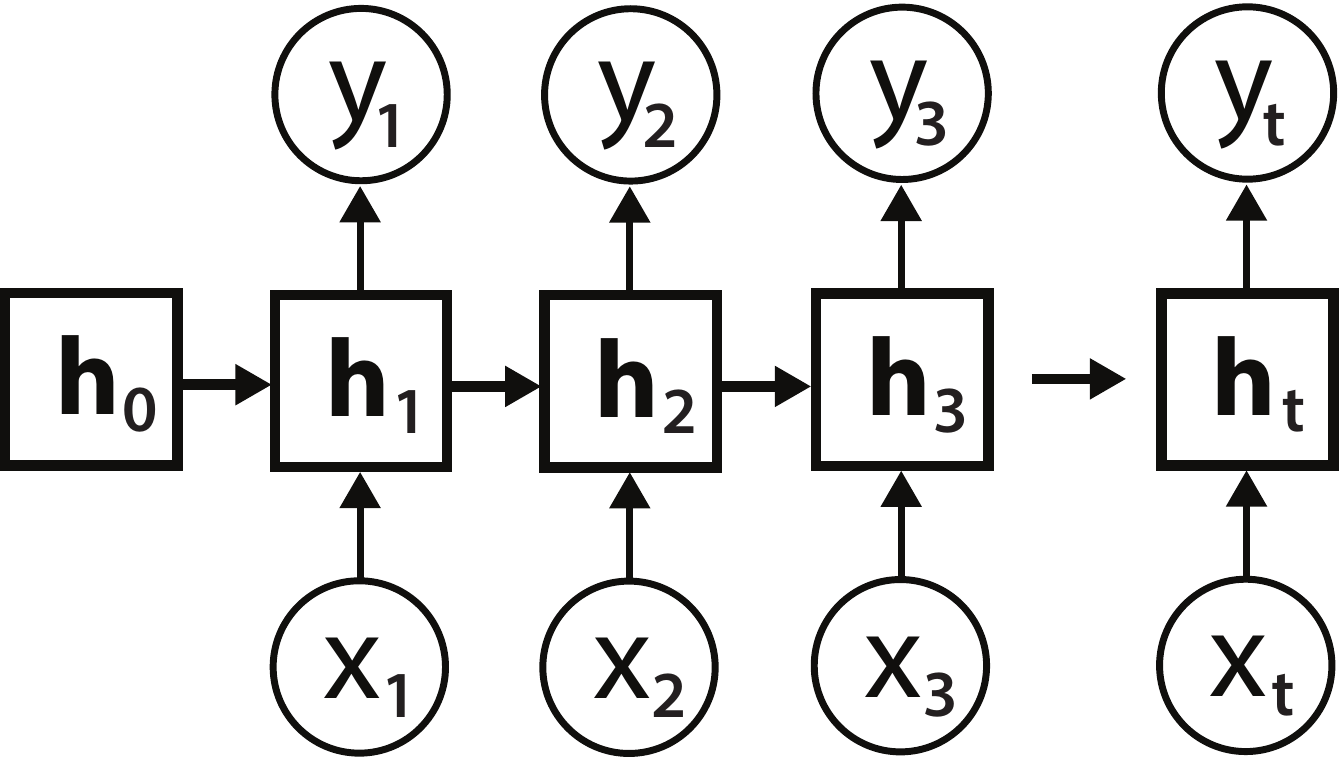}
\caption{Simple recurrent neural network}
\label{fig:rnn}
\end{figure}

Recurrent neural networks (RNNs) are a family of networks that can connect neurons over time, meaning that sequences of arbitrary length can be fed into RNNs. Crucially, RNNs incorporate a high dimensional, continuous latent state. This representation allows RNNs to use information from the past to impact a prediction at a later point in time. In this work, each input into the RNN will be a granular student action from a MOOC dataset. The RNN is trained to predict the student's next action. Figure \ref{fig:rnn} shows a diagram of a simple recurrent neural network, where inputs would be student actions and outputs would be the next student action from the sequence. 

\begin{align}
\boldsymbol{h}_t &= tanh(\boldsymbol{W}_{x}\boldsymbol{x}_t + \boldsymbol{W}_{h}\boldsymbol{h}_{t-1} + \boldsymbol{b}_h)\\
\boldsymbol{y}_t &= \sigma(\boldsymbol{W}_{y}\boldsymbol{h}_t + \boldsymbol{b}_y)
\end{align}

The RNN model is parameterized by an input weight matrix $\boldsymbol{W}_{x}$, recurrent weight matrix $\boldsymbol{W}_{h}$, initial state $\boldsymbol{h}_0$, and output matrix $\boldsymbol{W}_{y}$. $\boldsymbol{b}_h$ and $\boldsymbol{b}_y$ are biases for latent and output units, respectively.

\subsection{LSTM Models}

A popular variant of the RNN is the Long Short-Term Memory \cite{lstm1997} architecture, which is thought to help RNNs train by the addition of "gates" that learn when to retain meaningful information in the latent state and when to clear or "forget" the latent state, allowing for meaningful long-term interactions to persist. LSTMs add additional gating parameters that are explicitly learned in order to determine when to clear and when to augment the latent state with useful information. Each hidden state $h_i$ is instead replaced by an LSTM cell unit, which contains additional gating parameters.  As a result of these gates, LSTMs have been found to train more effectively than simple RNNs \cite{bengio1994learning, gers2000learning}. The update equations for an LSTM are:

\begin{align}
\boldsymbol{f}_t &= \sigma(\boldsymbol{W}_{fx}\boldsymbol{x}_t+\boldsymbol{W}_{fh}\boldsymbol{h}_{t-1} + \boldsymbol{b}_f)\\
\boldsymbol{i}_t &= \sigma(\boldsymbol{W}_{ix}\boldsymbol{x}_t + \boldsymbol{W}_{ih}\boldsymbol{h}_{t-1} + \boldsymbol{b}_i) \\
\boldsymbol{\tilde{C}}_t &= tanh(\boldsymbol{W}_{Cx}\boldsymbol{x}_t + \boldsymbol{W}_{Ch}\boldsymbol{h}_{t-1} + \boldsymbol{b}_C) \\
\boldsymbol{C}_t &= \boldsymbol{f}_t \times \boldsymbol{C}_{t-1} + \boldsymbol{i}_t \times \boldsymbol{\tilde{C}}_t\\
\boldsymbol{o}_t &= \sigma(\boldsymbol{W}_{ox}\boldsymbol{x}_t+\boldsymbol{W}_{oh}\boldsymbol{h}_{t-1} + \boldsymbol{b}_o)\\
\boldsymbol{h}_t &= \boldsymbol{o}_t \times tanh(\boldsymbol{C}_t)
\end{align}

\begin{figure}[!h]
\centering
\includegraphics[width=80mm]{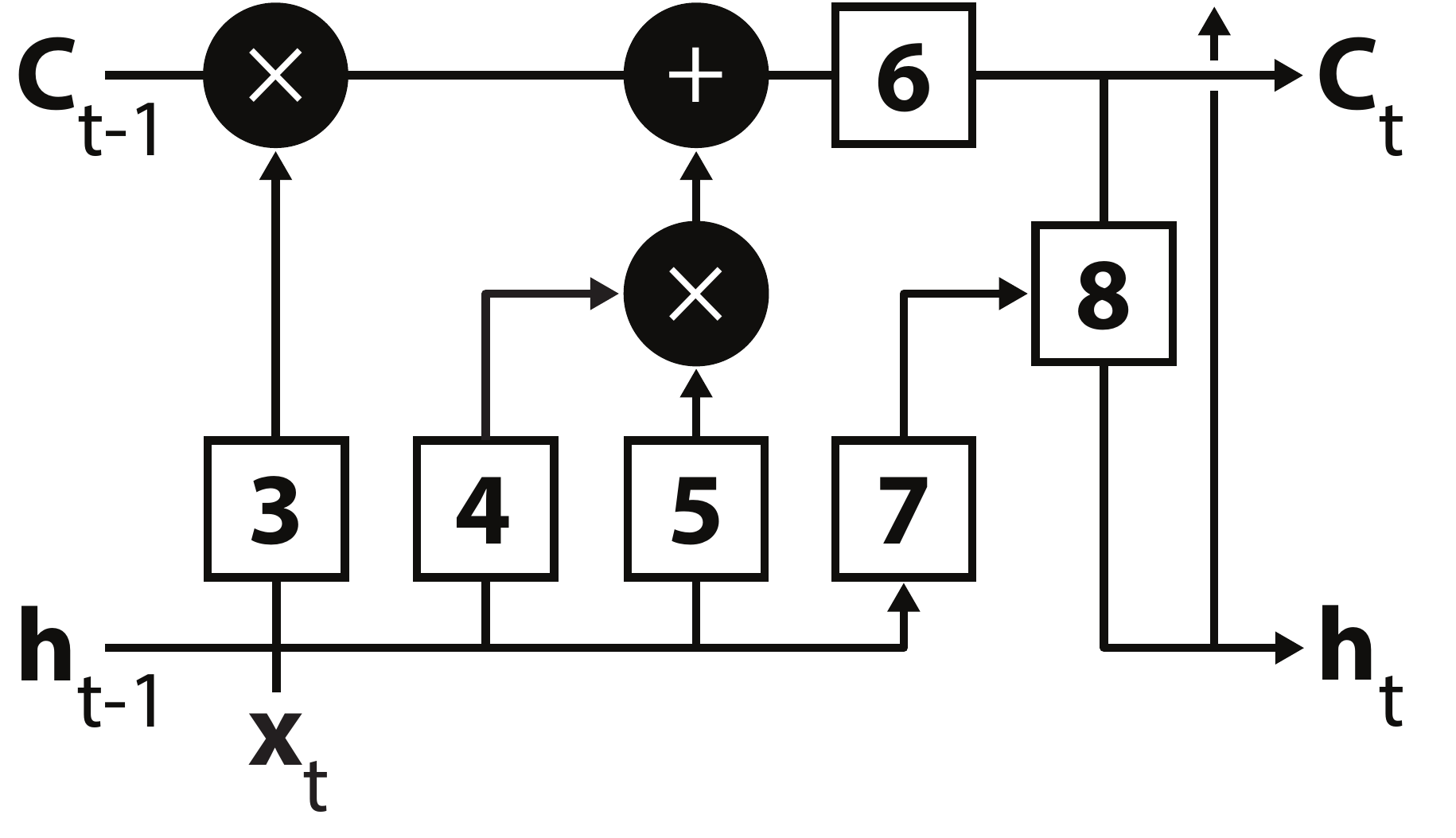}
\caption{An LSTM cell, where the numbers correspond to previously mentioned update equations}
\label{lstmcell}
\end{figure}

Figure \ref{lstmcell} illustrates the anatomy of a cell, where the numbers in the figure correspond to the previously mentioned update equations for the LSTM. $\boldsymbol{f}_t$, $\boldsymbol{i}_t$, and $\boldsymbol{o}_t$ represent the gating mechanisms used by the LSTM to determine ``forgetting'' data from the previous cell state, what to ``input'' into the new cell state, and what to output from the cell state. $\boldsymbol{C}_t$ represents the latent cell state for which information is removed from and added to as new inputs are fed into the LSTM. $\boldsymbol{\tilde{C}}_t$ represents an intermediary new candidate cell state that is gated to update the next cell state. 

\subsubsection{LSTM Implementation}
The generative LSTM models used in this paper were implemented using Keras \cite{Keras}, a Python library built on top of Theano \cite{bergstra+al:2010-scipy,Bastien-Theano-2012}. The model takes each student action represented by an index number. These indices correspond to the index in a 1-hot encoding of vectors, also known as dummy variabilization. The model converts each index to an embedding vector, and then consumes the embedded vector one at a time. The use of an embedding layer is common in natural language processing and language modelling \cite{DBLP:journals/corr/GoldbergL14} as a way to map words to a multi dimensional semantic space. An embedding layer is used here with the hypothesis that a similar mapping may occur for actions in the MOOC action space. The model is trained to predict the next student action, given actions previously taken by the student. Back propagation through time \cite{werbos1988generalization} is used to train the LSTM parameters, using a softmax layer with the index of the next action as the ground truth. Categorical cross entropy is used calculating loss, and RMSprop is used as the optimizer. Drop out layers were added between LSTM layers as a method to curb overfitting \cite{pham2014dropout}. Drop out randomly zeros out a set percentage of network edge weights for each batch of training data. In future work, it may be worthwhile to evaluate other regularization techniques crafted specifically for LSTMs and RNNs \cite{zaremba2014recurrent}.

\subsubsection{LSTM Hyperparameter Search}
As part of our initial investigation, we trained a total of 21 LSTM models for 10 epochs each. The searched space of hyperparameters for our LSTM models are shown in Table \ref{table:params}. These hyperparameters were chosen for grid search based on previous work which prioretized different hyperparameters based on effect size \cite{greff2015lstm}. For the sake of time, we chose not to train 3-layer LSTM models with learning rates of .0001.

We also performed an extended investigation, where we used the results from the initial investigation to serve as a starting point to explore additional hyperparameter and training methods. Because training RNNs is relatively time consuming, the extended investigation consisted of a subset of promising hyperparameter combinations. Results of the extended investigation are included in the Results section.

\begin{table}
\centering
\caption{LSTM Hyperparameter Grid}
\label{table:params}
\vspace{8 pt}

\begin{tabular}{l|ccc}
Hidden Layers & 1 & 2 & 3 \\
Nodes in Hidden Layer & 64 & 128 & 256 \\     
Learning Rate $(\eta)$ & .01 & .001 & .0001* \\
\end{tabular}
\end{table}

\subsubsection{Cross Validation}
To evaluate the predictive power of each model, 5-fold cross validation was used. Each model was trained on 80\% of the data and then validated on a remaining 20\%; this was done five times so that each set of student actions was in a validation set once. For the LSTMs, the model held out 10\% of its training data to serve as the hill climbing set to provide information about validation accuracy during the training process.  Each row in the held out set consists of the entire sequence of actions a student took. The proportion of correct next action predictions produced by the model is computed for each sequence of student actions. The proportions for an entire fold are averaged together to generate the model's performance for that particular fold, and then the performances across all 5 folds are averaged together to generate the \emph{CV-accuracy} for a particular LSTM model hyperparameter set.

\subsection{Shallow Models}

\subsubsection{N-gram Model}
\textit{N}-gram models are simple, yet powerful probabilistic models that aim to capture the structure of sequences through the statistics of \textit{n}-sized sub-sequences called \textit{grams} and are equivalent to n-order Markov Chains. Specifically, the model predicts each sequence state $x_{i}$ using the estimated conditional probability $P(x_{i} | x_{i-(n-1)},...,x_{i-1})$, which is the probability that $x_{i}$ follows the previous \textit{n}-1 states in the training set. \textit{N}-gram models are both fast and simple to compute, and have a straightforward interpretation. We expect \textit{n}-grams to be an extremely competitive standard, as they are relatively high parameter models that essentially assign a parameter per possible action in the action space. 

\subsubsection{N-gram Model Structure}
For the \textit{n}-gram models, we evaluated models where \textit{n} ranged from 2 to 10, the largest of which corresponds to the size of our LSTM context window during training. To handle predictions in which the training set contained no observations, we employed \textit{backoff}, a method that recursively falls back on the prediction of the largest \textit{n}-gram that contains at least one observation. Our validation strategy was identical to the LSTM models, wherein the average cross-validation score of the same five folds was computed for each model.

\subsubsection{Course Structure Models}

We also included a number of alternative models aimed at exploiting hypothesized structural characteristics of the sequence data. The first thing we noticed when inspecting the sequences was that certain actions are repeated several times in a row. For this reason, it is important to know how well this assumption alone predicts the next action in the dataset. Next, since course content is most often organized in a fixed sequence, we evaluated the the ability of the course syllabus to predict the next page or action. We accomplished this by mapping course content pages to student page transitions in our action set, which yielded an overlap of 174 matches out of the total 300 items in the syllabus. Since we relied on matching content ID strings that were not always present in our action space, a small subset of possible overlapping actions were not mapped. Finally, we combined both models, wherein the current state was predicted as the next state if the current state was not in the syllabus.

\section{Results}
In this section, we discuss the results from the previously mentioned LSTM models trained with different learning rates, number of hidden nodes per layer, and number of LSTM layers. Model success is determined through 5-fold cross validation and is related to how well the model predicts the next action. N-gram models, as well as other course structure models, are validated through 5-fold cross validation.

\subsection{LSTM Models}

Table \ref{table:paramperformance} shows the CV-accuracy for all 21 LSTM models computed after 10 iterations of training. For the models with learning rate of .01, accuracy on the hill climbing sets generally peaked at iteration 10. For the models with the lower learning rates, it would be reasonable to expect that peak CV-accuracies would improve through more training. We chose to simply report results after 10 iterations instead to provide a snapshot of how well these models are performing during the training process. We also hypothesize that model performance is unlikely to improve drastically over the .01 learning rate model performances in the long run, and we have the need to maximize the most promising explorations to run on limited GPU computation resources. The best CV-accuracy for each learning rate is bolded for emphasis.

\begin{table}[!h]
\centering
\caption{LSTM Performance (10 Epochs)}
\vspace{8 pt}
\label{table:paramperformance}
\begin{tabular}{lccc}
\hline
\noalign{\vskip 1pt}
Learn Rate    & Nodes  & Layers & Accuracy   \\ 
\noalign{\vskip 1pt}
\hline
\noalign{\vskip 2pt}
0.01   & 64  & 1  & 0.7014 \\ 
0.01   & 64  & 2  & 0.7009 \\ 
0.01   & 64  & 3  & 0.6997 \\
0.01   & 128 & 1  & 0.7046 \\ 
0.01   & 128 & 2  & 0.7064 \\ 
0.01   & 128 & 3  & 0.7056 \\
0.01   & 256 & 1  & 0.7073 \\ 
0.01   & 256 & 2  & \textbf{0.7093} \\
0.01   & 256 & 3  & 0.7092 \\
\noalign{\vskip 1pt}
\hline
\noalign{\vskip 2pt}
0.001  & 64  & 1  & 0.6941 \\ 
0.001  & 64  & 2  & 0.6968 \\ 
0.001  & 64  & 3  & 0.6971 \\ 
0.001  & 128 & 1  & 0.6994 \\ 
0.001  & 128 & 2  & 0.7022 \\ 
0.001  & 128 & 3  & 0.7026 \\ 
0.001  & 256 & 1  & 0.7004 \\ 
0.001  & 256 & 2  & \textbf{0.7050} \\ 
0.001  & 256 & 3  & \textbf{0.7050} \\ 
\noalign{\vskip 1pt}
\hline
\noalign{\vskip 2pt}
0.0001 & 64  & 1  & 0.6401 \\ 
0.0001 & 64  & 2  & 0.4719 \\ 
0.0001 & 128 & 1  & 0.6539 \\ 
0.0001 & 128 & 2  & 0.6648 \\ 
0.0001 & 256 & 1  & 0.6677 \\ 
0.0001 & 256 & 2  & \textbf{0.6894} \\ 
\end{tabular}
\end{table}

One downside of using LSTMs is that they require the use of a GPU and are relatively slow to train. Thus, when investigating the best hyperparameters to use, we chose to train additional models based only on a subset of the initial explorations. We also extend the amount of context exposed to the model, extending past context from 10 elements to 100 elements. Table \ref{table:extendperformance} shows these extended results. Each LSTM layer has 256 nodes and is trained for either 20 or 60 epochs, as opposed to just 10 epochs in the previous hyperparameter search results. The extended results show a large improvement over the previous results, where the new accuracy peaked at .7223 compared to .7093.

\begin{table}[!h]
\centering
\caption{Extended LSTM Performance (256 Nodes, 100 Window Size)}
\vspace{8 pt}
\label{table:extendperformance}
\begin{tabular}{lccc}
\hline
\noalign{\vskip 1pt}
Learn Rate    & Epochs  & Layers & Accuracy   \\ 
\noalign{\vskip 1pt}
\hline
\noalign{\vskip 2pt}
0.01   & 20  & 2  & 0.7190 \\ 
0.01   & 60  & 2  & 0.7220 \\ 
0.01   & 20  & 3  & 0.7174 \\
0.01   & 60  & 3  & \textbf{0.7223} \\ 
\noalign{\vskip 1pt}
\hline
\noalign{\vskip 2pt}
0.001  & 20  & 2  & 0.7044 \\ 
0.001  & 60  & 2  & 0.7145 \\ 
0.001  & 20  & 3  & 0.7039 \\ 
0.001  & 60  & 3  & \textbf{0.7147} \\ 
\end{tabular}
\end{table}

\subsubsection{Performance During Training}
\begin{figure*}
\centering
\includegraphics[width=\linewidth,keepaspectratio]{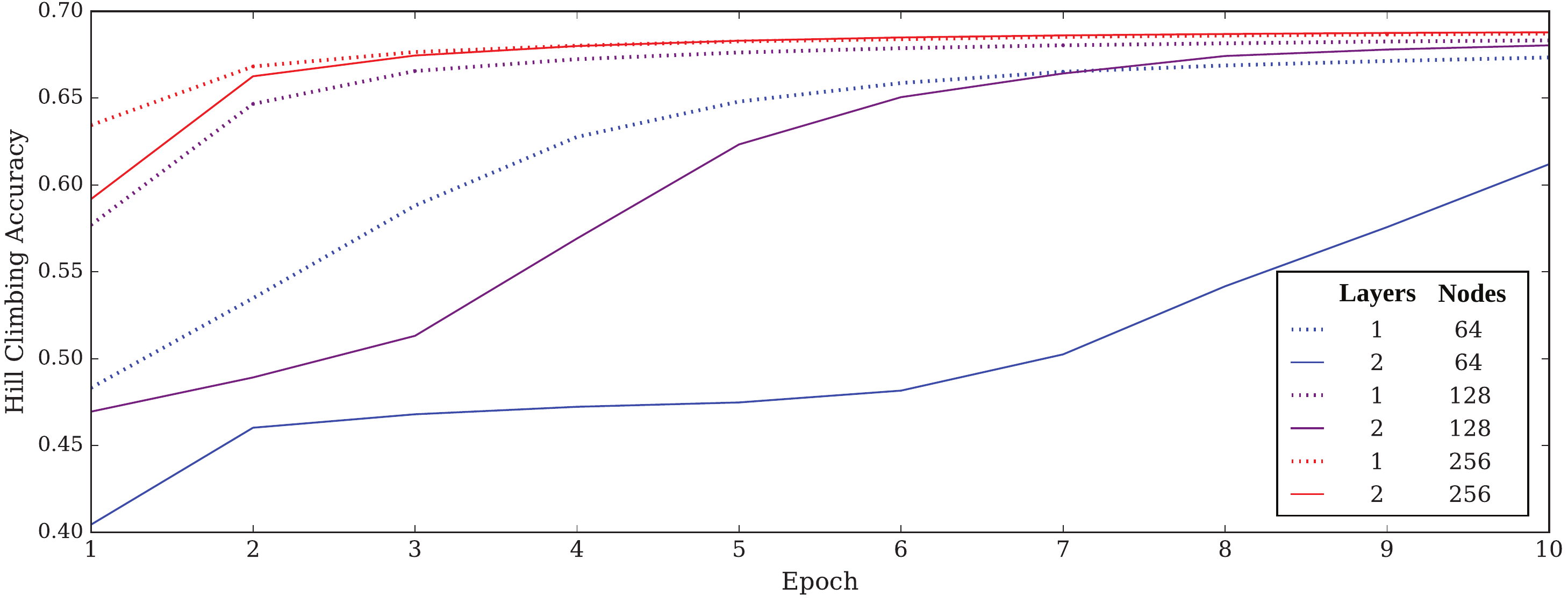}
\caption{Average accuracy by epoch on hill climbing  data which comprised 10\% of each training set} 
\label{perform-by-epoch}
\end{figure*}

Figure \ref{perform-by-epoch} shows validation accuracy on the 10\% hill-climbing hold out set during training by epoch for the 1 and 2 layer models from the initial exploration. Each data point represents the average hill-climbing accuracy among all three learning rates for a particular layer and node count combination. Empirically, having a higher number of nodes is associated with a higher accuracy in the first 10 epochs, while 2 layer models start with lower validation accuracies for a few epochs before approaching or surpassing the corresponding 1 layer model. This figure provides a snapshot for the first 10 epochs; it is clearly the case that for some parameter combinations, more epochs would result in a higher hill-climbing accuracy, as shown by the additional extended LSTM search. Extrapolating, 3-layer models may also follow the trend that the 2-layer models exhibited where accuracies may start lower initially before improving over their lower layer counterparts.

\subsection{Course Structure Models}

Model performance for the different course structure models is shown in Table \ref{table:structure-perform}. Results suggest that many actions can be predicted from simple heuristics such as stationarity (same as last), or course content structure. Combining both of these heuristics ("syllabus + repeat") yields the best results, although none of the alternative models obtained performance within the range of the LSTM or \textit{n}-gram results. 


\begin{table}[!h]
\centering
\caption{Structural Models}
\vspace{8 pt}
\label{table:structure-perform}
\begin{tabular}{lc}
\hline
Structural Model  &   Accuracy \\
\noalign{\vskip 1pt}
\hline
\noalign{\vskip 2pt}
repeat  &  0.2908 \\
syllabus  &  0.2339 \\
syllabus + repeat  &  \textbf{0.4533} \\
\end{tabular}
\end{table}

\subsection{N-gram Models}
\begin{table}[!h]
\centering
\caption{N-gram Performance}
\vspace{8 pt}
\label{table:ngram-perform}
\begin{tabular}{lc}
\hline
\noalign{\vskip 1pt}
\textit{N}-gram  &   Accuracy   \\
\noalign{\vskip 1pt}
\hline
\noalign{\vskip 2pt}
2-gram  &  0.6304 \\
3-gram  &  0.6658 \\
4-gram  &  0.6893 \\
5-gram  &  0.6969 \\
6-gram  &  0.7012 \\
7-gram  &  0.7030 \\
8-gram  &  \textbf{0.7035} \\
9-gram  &  \textbf{0.7035} \\
10-gram &  0.7033 \vspace{5pt} \\
\end{tabular}
\end{table}

Model performance is shown in Table \ref{table:ngram-perform}. The best performing models made predictions using either the previous 7 or 8 actions (8-gram and 9-gram respectively). Larger histories did not improve performance, indicating that our range of \textit{n} was sufficiently large. Performance in general suggests that \textit{n}-gram models were competitive with the LSTM models, although the best \textit{n}-gram model performed worse than the best LSTM models. Table \ref{table:ngram-prop} shows the proportion of n-gram models used for the most complex model (10-gram). More than $62\%$ of the predictions were made using 10-gram observations. Further, less than $1\%$ of cases fell back on unigrams or bigrams to make predictions, suggesting that there was not a significant lack of observations for larger gram patterns. Still,  about 6\% fewer data points looks to be predicted by successively larger n-grams.

\begin{table}[!hb]
\centering
\caption{Proportion of 10-gram prediction by \textit{n}}
\vspace{8 pt}
\label{table:ngram-prop}
\begin{tabular}{cc}
\hline
\noalign{\vskip 1pt}
\textit{n}  &  \% Predicted by \\
\noalign{\vskip 1pt}
\hline
\noalign{\vskip 2pt}
1  & 0.0003 \\
2  & 0.0084 \\
3  & 0.0210 \\
4  & 0.0423 \\
5  & 0.0524 \\
6  & 0.0605 \\
7  & 0.0624 \\
8  & 0.0615 \\
9  & 0.0594 \\
10 & \textbf{0.6229} \\
\end{tabular}
\end{table}

\subsection{Validating on Uncertified Students}
We used the best performing ``original'' LSTM model after 10 epochs of training (.01 learn rate, 256 nodes, 2 layers) to predict actions on streams of data from students who did not ultimately end up certified. Many uncertified students only had a few logged actions, so we restricted analysis to students who had at least 30 logged actions. There were 10761 students who met this criteria, with a total of 2151662 actions. The LSTM model was able to correctly predict actions from the uncertified student space with .6709 accuracy, compared to .7093 cross validated accuracy for certified students. This difference shows that actions from certified students tend to be different than actions from uncertified students, perhaps showing potential application in providing an automated suggestion framework to help guide students.

\section{Contribution}

In this work, we approached the problem of modelling granular student action data by modelling all types of interactions within a MOOC. This differs in approach from previous work which primarily focuses on modelling latent student knowledge using assessment results. In predicting a student's next action, the best performing LSTM model produced a cross-validation accuracy of .7223, which was an improvement over the best \textit{n}-gram model accuracy of .7035. This amounts to $210,000$ more correct predictions of the total 11-million possible. Table \ref{table:cvmodels} shows the number of times the two models agreed or disagreed on a correct or an incorrect prediction during cross validation. Both LSTM and \textit{n}-gram models provide significant improvement over the structural model of predicting the next action by syllabus course structure and through repeats, which shows that there are patterns of student engagement that clearly deviate from a completely linear navigation through the course material. 

It's important to note that these results are the best LSTM models explored among the subset of hyperparameters detailed in this paper, and that additional improvements may be found with more investigation. The latent architecture of the LSTM model is clearly manifesting a performance boost in predictions of validation sets, perhaps showing that the model can use information from further back more effectively.

\begin{table}
\centering
\caption{Cross Validated Models Comparison}
\vspace{5pt}
\label{table:cvmodels}
\begin{tabular}{lcc}
 & N-gram Correct & N-gram Incorrect \vspace{1pt} \\ 
\multicolumn{1}{l|}{LSTM Correct} & 7,565,862         & 577,683         \\
\multicolumn{1}{l|}{LSTM Incorrect} & 367,960        & 2,735,702         \\ 
\end{tabular}
\end{table}

\section{Future Work}
\subsection{Model and Dataset Improvements}
Both the LSTM and the \textit{n}-gram models have room for improvement. In particular, our \textit{n}-gram models could benefit from a combination of backoff and smoothing techniques, which allow for better handling of unseen grams. Our LSTM may benefit from a broader hyperparameter grid search, more training time, longer training context windows, and higher-dimensional action embeddings. Additionally, the signal-to-noise ratio in our dataset could be increased by removing less informative or redundant student actions, or adding additional tokens to represent time between actions.

\subsection{Applications}
The primary reason for applying deep learning models to large sets of student action data is to model student behavior in MOOC settings, which leads to insights about how successful and unsuccessful students navigate through the course. These patterns can be leveraged to help in the creation of automated recommendation systems, wherein a struggling student can be provided with transition recommendations to view content based on their past behavior and performance. To evaluate the possiblity of such an application, we plan to experimentally test a recommendation system derived from our network against an undirected control group.

\subsection{Incorporating Larger Datasets}
In this paper, we examined student actions from a single MOOC course. Future work should assess performance of similar models for a variety of courses and examine to what extent course-general patterns can be learned using a single model.

\section*{Acknowledgement}
This work was supported by a grant from the National Science Foundation (IIS: BIGDATA 1547055).

\bibliographystyle{unsrt}
\bibliography{ACM_proc}


\end{document}